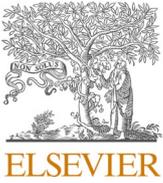
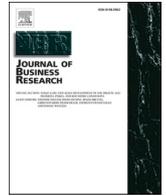

# Ecosystem orchestration practices for industrial firms: A qualitative *meta*-analysis, framework development and research agenda

Lei Shen [a], Qingyue Shi [a], Vinit Parida [b,c,*], Marin Jovanovic [d]

[a] *Glorious Sun School of Business and Management, Donghua University, Shanghai 200051, China*
[b] *Entrepreneurship and Innovation, Luleå University of Technology, 97187, Luleå, Sweden*
[c] *Department of Management, University of Vaasa, PO Box 700, FI-65101 Vaasa, Finland*
[d] *Department of Operations Management, Copenhagen Business School, DK-2000 Frederiksberg, Denmark*

A B S T R A C T

This study ventures into the dynamic realm of ecosystem orchestration for industrial firms, emphasizing its significance in maintaining competitive advantage in the digital era. The fragmented research on this important subject poses challenges for firms aiming to navigate and capitalize on ecosystem orchestration. To bridge this knowledge gap, we conducted a comprehensive qualitative *meta*-analysis of 31 case studies and identified multifaceted orchestration practices employed by industrial firms. The core contribution of this research is the illumination of five interdependent but interrelated orchestration practices: strategic design, relational, resource integration, technological, and innovation. Together, these practices are synthesized into an integrative framework termed the "Stirring Model," which serves as a practical guide to the orchestration practices. Furthermore, the conceptual framework clarifies the synergy between the identified practices and highlights their collective impact. This study proposes theoretical and practical implications for ecosystem orchestration literature and suggests avenues for further research.

## 1. Introduction

Over the past two decades, theoretical studies of ecosystems have become increasingly prevalent and valuable in strategic management research (Altman et al., 2022; Jacobides et al., 2018; Jacobides et al., 2024). Firms manage external environment dependencies through networks established by formal bi-lateral inter-organizational relationships, which are strategically significant (Shipilov & Gawer, 2020). Ecosystems, on the other hand, comprise a set of actors with varying degrees of multi-lateral, non-generic complementarities that are not fully hierarchically controlled (Jacobides et al., 2018). Therefore, ecosystems are characterized by greater openness, wider heterogeneity of partners, and ecosystem-level outputs that create value for the final consumer (Kapoor, 2018). Orchestration has emerged in both literature streams as a necessary management approach facilitating activities within such loosely coupled systems, networks (Dhanaraj & Parkhe, 2006; Giudici et al., 2018), and ecosystems (Autio, 2022; Sjödin et al., 2022). However, the hierarchically independent yet technologically and economically interdependent nature of multilateral ecosystem partners and the focal ecosystem value proposition make ecosystem orchestration a distinct strategic management concept (Autio, 2022; Thomas & Autio, 2020). Ecosystem orchestration involves deliberate, purposeful actions usually taken by a hub firm, a keystone player, or an orchestrator (Dhanaraj & Parkhe, 2006; Iansiti & Levien, 2004; Iyer et al., 2006). The literature argues that an ecosystem orchestrator "may play an indispensable role in supporting the ongoing recombination of members' knowledge to explore new opportunities" (Giudici et al., 2018). Although the role of ecosystem orchestrator is perceived as offering significant prospects for value capture within the ecosystem (Rietveld et al., 2020), the objective of orchestration is to create and appropriate value among all partners in the ecosystem (Altman et al., 2022; Dattée et al., 2018; Majchrzak et al., 2023).

The ecosystem orchestrator requires non-contractual orchestration practices (Autio, 2022). These practices are imperative for fostering voluntary, proactive co-creation among ecosystem partners (Autio, 2022). For instance, Reypens et al. (2021) identified three fundamental orchestration practices - connecting, facilitating, and governing - used by orchestrators to mobilize diverse ecosystem partners. Nevertheless, given the multifaceted nature of ecosystem orchestration, the existing research body remains comparatively segmented. First, orchestration pertains to various phases of ecosystem evolution, encompassing the spectrum from emergence (Daymond et al., 2022; Pushpananthan & Elmquist, 2022; Thomas et al., 2022; Gomes, Facin, et al., 2022) through expansion or momentum to the control phase (Autio, 2022; Holgersson et al., 2022; Kolagar et al., 2022). The notion of the orchestrating actor introduces additional complexity, as different ecosystem stages may require varied orchestrators, especially highlighted in the industrial firm context (Lingens, Miéhé, et al., 2021). For example, an ecosystem orchestrator may facilitate the emergence phase with specific enabling practices (Blackburn et al., 2022), while specialized orchestrators might be better equipped for subsequent phases, like commercializing the ecosystem's value proposition (Lingens, Seeholzer, et al., 2022). Second, orchestration can manifest in two forms: dominating orchestration, where orchestrators assume a visionary and directive role to shape ecosystem trajectories and interactions actively, and consensus-based orchestration, where orchestrators create a trust-based, adaptable environment allowing collaborative dynamics to steer ecosystem evolution organically (Reypens et al., 2021). The dominating approach for industrial firms tends to align with top-down practices, characterized by orchestrators imposing a preconceived structure and direction on the ecosystem. Conversely, the consensus-based approach correlates with bottom-up practices, where the ecosystem's value proposition evolves organically without being predefined (Autio, 2022). Finally, orchestration spans various domains, including architecture, community, governance (Thomas & Tee, 2022), and behavioral (Ritala et al., 2023), necessitating multifaceted practices for effective ecosystem growth management. Therefore, a comprehensive examination of orchestration practices is essential. Such exploration will provide insights into ecosystem orchestrators' practices to create and capture value within ecosystems, aiming to fulfill a collective vision (Jacobides et al., 2024). Understanding these practices and their dynamics is crucial for any ecosystem orchestrator aspiring to navigate and succeed with ecosystems in the industrial context.

In line with the above-stated research gaps, the purpose of this is twofold. First, identify and synthesize ecosystem practices for industrial firms, and second, develop propositions and research agenda for stimulating future research on ecosystem orchestration practices. We conducted a qualitative *meta*-analysis to consolidate existing research results on ecosystem orchestration for industrial firms. Our analysis draws insights from 31 case studies published before December 2022, revealing the critical role of five distinct ecosystem orchestration practices: strategic design, relational, resource integration, technological leveraging, and innovation practices. These practices encompass 15 key orchestration activities and 30 sub-activities, culminating in the developing of an integrated framework named the "Stirring Model". This study pioneers the application of qualitative *meta*-analysis in the field of ecosystem orchestration. By integrating perspectives from multiple industries, our research presents a comprehensive framework for ecosystem orchestration practices with high external validity, serving as a guidebook and assessment tool for ecosystem orchestrators and partners.

The rest of the paper is structured as follows: Section 2 elaborates on the theoretical background and reviews relevant literature. Section 3 covers research methodology, data sources, and analysis. The subsequent section introduces a multilayered perspective to ecosystem orchestration practices, develops our framework, and formulates eight propositions. Section 5 discusses theoretical contributions, managerial implications, limitations, and future directions.

## 2 Theoretical background

### 2.1. Understanding ecosystem concept

In recent decades, the term "ecosystem" has gained prominence in both research and practice (Adner, 2017; Altman et al., 2022; Jacobides et al., 2018). This concept offers a novel perspective on organizational collective action and understanding of the competitive environment (Thomas & Ritala, 2022). These advancements highlight a growing interest in exploring interdependencies across organizations (Adner, 2017) and in the formation of *meta*-organizations (Gulati et al., 2012; Kretschmer et al., 2022). Owing to the dispersed nature of co-specialized knowledge and the network effects from technological platforms (Gawer, 2021; Kretschmer et al., 2022), firms increasingly find themselves limited to creating and capturing value within their boundaries (Ritala & Jovanovic, 2023). The nature of co-specialized knowledge and its complementary effects have rendered traditional markets inefficient (Foss et al., 2023), necessitating structures for co-creation and co-production practices within the ecosystem domain (Ceccagnoli et al., 2012). Consequently, firms are compelled to engage in collaborations within ecosystems (Jacobides et al., 2024).

The origins of ecosystem research can be traced back to Moore (1993), who initially presented the ecosystem construct as a metaphor borrowed from biology. In recent years, the term "ecosystem" has evolved to encompass a variety of meanings. Adner (2017) categorized these interpretations into two perspectives: (a) ecosystem-as-affiliation, which considers ecosystems as communities of associated actors defined by their networks and platform affiliations, and (b) ecosystem-as-structure, which views ecosystems as configurations of activity defined by a value proposition. Recently, there has been emerging consensus on the notion of an ecosystem as a structure of interdependence (Lingens, Miéhé, et al., 2021; Gomes, dos Santos, et al., 2022; Gomes, Facin, et al., 2022). Aligning with this prevailing perspective, we adopt the definition of an ecosystem as "an alignment structure of interdependent but hierarchically independent heterogeneous partners collectively striving to materialize a shared value proposition" (Adner, 2017; Autio, 2022; Thomas & Ritala, 2022). This definition underscores the shared value proposition as the cornerstone of collaborative arrangements, incentivizing actors to collaborate and invest in co-specialized assets and activities for joint value creation (Jacobides et al., 2024). It also plays a crucial role in shaping ecosystem boundaries (Adner, 2017; Thomas & Ritala, 2022). Moreover, the autonomy of ecosystem partners necessitates seeking non-hierarchical ways to orchestrate ecosystems (Autio, 2022). For instance, the alignment structure may also involve mutual agreements on standards for interoperability and respective business models within the integrated bundle of the shared value proposition (Kapoor, 2018; Kohtamäki et al., 2019; Wareham et al., 2014). Furthermore, based on the functions performed by ecosystem partners, the ecosystem aligns with firms whose modules are mutually complementary and unique, indicating a state of interdependence where firms add value to each other and are not easily substitutable (Jacobides et al., 2018; Lingens, Böger, et al., 2021). Essentially, ecosystems cannot be reduced to bi-lateral relationships (Adner, 2017; Jacobides et al., 2018). Ecosystem partners collectively deliver outputs that surpass what any single firm could achieve in isolation (Adner, 2017; Lingens, Miéhé, et al., 2021; Thomas & Ritala, 2022).

## 2.2. Ecosystem orchestration practices and activities

Orchestration[1] is defined as "a set of deliberate, purposeful actions" taken by an orchestrator (Dhanaraj & Parkhe, 2006) aimed at unlocking the potential and coordinating the efforts of other firms while fostering a common vision through organizational and social norms (Carida et al., 2022). This term has become prominent in strategic management and marketing literatures, especially when describing the actions of a focal firm that facilitate value-creation interactions among various actors, moving beyond the traditional supplier-customer relationship (Mann et al., 2022; Shi & Shen, 2022). Its empowering nature has secured orchestration a central position in ecosystem literature (Autio, 2022; Lingens, Böger, et al., 2021). In the context of an industrial 5G ecosystem, effective orchestration involves the collaboration of telecommunications companies, hardware manufacturers, software developers, industrial enterprises, regulatory bodies, cybersecurity experts, research institutions, logistics partners, and consulting services, each playing a distinct yet interconnected role in implementing and optimizing 5G technology in smart manufacturing environments. Consequently, orchestration has emerged as a dominant theme in the industrial ecosystem literature (Pattinson et al., 2022), attracting significant attention from both academic and business communities.

Drawing on existing insights, we aim to distinguish and synthesize two general perspectives: (a) ecosystem orchestration-as-ability and (b) ecosystem orchestration-as-activities. Orchestration-as-ability conceptualizes ecosystem orchestration as a combination of capabilities that enable and drive ecosystem development. Pitelis and Teece (2018) argue that orchestration theory is grounded in and encompasses the dynamic capabilities framework. For instance, Schreieck et al. (2021) view ecosystem orchestration as a relationship-driven capability that assists the orchestrator in enabling and balancing value co-creation and capture within an emergent ecosystem. Foss et al. (2023) emphasize ecosystem leadership that is facilitating the formation of a shared vision (sensing), inducing others to make ecosystem-specific investments (seizing), and engaging in ad-hoc problem solving to create and maintain stability (reconfiguring/transforming). In contrast, orchestration-as-activities involves a dynamic set of evolving actions by ecosystem orchestrators to leverage complementary actors for a joint value proposition (Lingens, Miehé, et al., 2021). Cennamo et al. (2022) define ecosystem orchestration as a series of processes that facilitate partner activities to enhance the ecosystem's value. Nambisan and Sawhney (2011) present orchestration processes of managing innovation leverage, managing innovation coherence, and managing innovation appropriability. Viewing orchestration as either capabilities or activities, the common focus remains on "what ecosystem orchestrators do". Therefore, we integrate these perspectives to investigate how ecosystem orchestrators can effectively manage ecosystems. This integration leads to a more comprehensive understanding of ecosystem orchestration.

Operating within ecosystems presents significant challenges due to tensions in value creation and capture (Cennamo & Santaló, 2019) or between cooperation and competition among ecosystem partners (Hannah & Eisenhardt, 2018; Kapoor & Lee, 2013). Therefore, ecosystems require a degree of control to ensure alignment towards the common value proposition (Jacobides et al., 2018; Lingens, Miehé, et al., 2021). This alignment is facilitated by a central actor known as the ecosystem orchestrator, who plays a pivotal role in balancing generativity and coherence in ecosystem outputs (Thomas & Ritala, 2022). Therefore, irrespective of the perspective on orchestration, the orchestrator remains a critical entity in managing and aligning the activities of the various actors within the ecosystem (B. Lingens, Huber, et al., 2022; Lingens, Miehé, et al., 2021).

## 2.3. Ecosystem orchestrators

An ecosystem orchestrator, often referred to as the hub or leader firm in ecosystems (Dhanaraj & Parkhe, 2006), leverages the benefits of a central position, sometimes termed as "smart power" (Williamson & De Meyer, 2012). This role enables orchestrators to act as brokers, influencing, mediating, or modifying relationships within the ecosystem (Halevy et al., 2019; Ritala et al., 2023). Therefore, the presence of an ecosystem orchestrator is a fundamental and prominent feature in an ecosystem (Linde et al., 2021; Sjödin et al., 2022). Notably, ecosystems with at least one orchestrator firm practicing effective orchestration tend to incur lower transaction costs compared to those without such leadership (Foss et al., 2023). Orchestrators are indispensable in aligning ecosystem partners towards a shared value proposition (Adner, 2017). They aim to steer the ecosystem toward unique value propositions that are difficult to replicate and imitate, relying on innovative business models, opportunity sharing, and the purposeful cultivation of long-term partnerships (Ritala et al., 2013).

However, not every company is capable of performing as an ecosystem orchestrator (Parida et al., 2019). The roles played by ecosystem orchestrators can be divided into market designers and market explorers (Isckia et al., 2020). As market designers, they act as architects of the emerging ecosystem, typically defining complementarity (Lingens, Seeholzer, et al., 2022). Orchestrators set system-level goals, define hierarchical distinctions among member roles, and establish standards and interfaces (Holgersson et al., 2022; Sjödin et al., 2022). Ecosystem development is driven by a partially structured process led by the orchestrator (Lingens, Seeholzer, et al., 2022). Additionally, orchestrators need to transition from an inward ("ego-centric") to an outward ("eco-centric") mindset (Mann et al., 2022; Shi et al., 2023) and strive to prolong their orchestration role (Leten et al., 2013). As market explorers, orchestrators function as environmental scanners to identify, evaluate, and capitalize on new business opportunities (Addo, 2022; Giudici et al., 2018). They actively seek new value creation opportunities (Kindermann et al., 2022), respond to business challenges (Reypens et al., 2021), alleviate ecosystem bottlenecks (Masucci et al., 2020), and continuously adapt to internal and external dynamic changes (Mann et al., 2022).

Ecosystem orchestration presents numerous challenges for orchestrators, predominantly in resource and relationship management. Orchestrators must possess valuable resources, such as unique technology (Gawer & Cusumano, 2014), provide a stable set of common assets (Parida et al., 2019), and align internal and external resources to create new market opportunities (Isckia et al., 2020; Mann et al., 2022; Zeng et al., 2021). They are responsible for orchestrating the flow of resources (Linde et al., 2021). Regarding relationships, these in the ecosystem extend beyond traditional commercial alliances, and their harmony heavily depends on intangible factors not outlined in formal agreements. Orchestrators need to manage relatively autonomous actors (Ge & Liu, 2022) and address the complexities of dealing simultaneously and temporally with a large and diverse number of partners (Reypens et al., 2021). They also have to integrate the reactions of other actors into their decision-making processes (Isckia et al., 2020). Additionally, the ecosystem orchestrator must safeguard not only its internal reconfiguring activities but also those of the ecosystem partners (Leten et al., 2013; Linde et al., 2021). This requires providing stability, displaying compliance, and ensuring conformity (Autio, 2022; Lingens, Huber, et al., 2022). While orchestrators have traditionally been large, established firms, recent research shows that small firms, startups, and even marginal firms can assume the role of orchestrators (Cui et al., 2019; Lingens, Miéhé, et al., 2021). Moreover, the number of orchestrators within an ecosystem can vary, leading to single, double, or multiple orchestration systems (Lingens, Huber, et al., 2022).

---

[1] In the study of ecosystems, it is important to distinguish between similar but distinct concepts like ecosystem governance, coordination, and orchestration. Ecosystem governance refers to a set of regulatory rules, including those for value capture, technological standards, and membership, which are designed to control the behavior of ecosystem members (Gomes et al., 2021). This governance typically relies on formal, hierarchical one-to-one supplier contracts (Autio, 2022), as opposed to ecosystem orchestration, which involves engaging partners voluntarily (Carida et al., 2022) and includes more situationally flexible activities. Ecosystem coordination, on the other hand, deals with breaking down, recording, and altering tasks, thereby creating new work formats for the coordinating body (Mann et al., 2022). Unlike orchestration, which aims to unite independent players for mutual benefit and a shared vision, coordination is primarily focused on benefiting the entity that coordinates these efforts (Mann et al., 2022).

Based on the level of control exerted by orchestrators, ecosystem orchestration modes can be classified into three categories: dominating, consensus-based, and hybrid orchestration. Dominating orchestration is a top-down approach where orchestrators exercise significant control and design influence, intentionally shaping the ecosystem's architecture (Autio, 2022; Reypens et al., 2021). In contrast, consensus-based orchestration operates on a bottom-up approach, prioritizing ongoing collaboration, negotiations, and voluntary engagement (Autio, 2022; Reypens et al., 2021), emphasizing collective action among different ecosystem participants (Thomas & Ritala, 2022). The third mode, hybrid orchestration, as proposed by Reypens et al. (2021), involves orchestrators switching between dominating and consensus-based modes in response to ecosystem challenges. We adopt this perspective, recognizing the need for adaptability in different situations.

The insights discussed here are primarily sourced from a range of qualitative studies, emphasizing case study methodologies. However, the heterogeneity in research foci, theoretical frameworks, and contexts of analysis has led to a lack of systematic organization and integration among the findings. Despite this, it is noteworthy that many orchestration practices demonstrate notable similarities across diverse industries, as highlighted by Ritala et al. (2013). This observation underscores the potential for employing qualitative *meta*-analysis as a method to amalgamate and refine these disparate results into a more cohesive understanding. Considering the pivotal role of ecosystem orchestrators, this study adopts an orchestrator-centric lens. Consequently, the research question posited is: What practices do ecosystem orchestrators employ to orchestrate their ecosystems effectively?

**3 Research methods**

*3.1. Qualitative meta-analysis*

This study employed a qualitative *meta*-analysis approach to address the research questions. This method was first introduced by Hoon in 2013 as "*meta*-synthesis of qualitative case studies" in the field of organization and management research, where he systematically explored its implications and operational steps (Hoon, 2013). In line with recent studies (e.g., Laubengaier et al. (2022); Küberling-Jost (2021)), we refer to this method as qualitative *meta*-analysis. It serves as an exploratory research design aimed at reinterpreting and synthesizing primary qualitative case studies to refine existing theories, extend them and even to generate new ones (Habersang et al., 2019). The primary objective of qualitative *meta*-analysis is to develop novel theoretical interpretations that surpass the findings reported in primary case studies (Combs et al., 2019). Case studies are particularly valuable as they usually explore the reasons and nature of events in particular contexts, providing rich and unique empirical descriptions that may often not fully captured by quantitative studies (Habersang et al., 2019; Rauch et al., 2014). However, due to the emphasis on the novelty in the research findings, different conclusions may arise for the same phenomenon (Hoon, 2013). To address this, the qualitative *meta*-analysis approach rigorously analyzes a large number of case studies, which can reconcile previous divergent and irreconcilable empirical evidence to provide more robust, generalizable, and comprehensive conclusions (Habersang et al., 2019). This is where qualitative *meta*-analysis differs greatly from literature review and particularly bibliometric study.

While literature review encompasses both qualitative and quantitative research, qualitative *meta*-analysis focuses exclusively on literature that employs the case study methodology. Additionally, literature review aims to integrate the findings of prior studies (Berente et al., 2019), thereby objectively identifying major contributors, topics, and research gaps in a field (Mukherjee et al., 2022). On the other hand, qualitative *meta*-analysis attempts to synthesize key variables and underlying relationships from published qualitative case studies to build a theory and go beyond the contributions in the original studies (Hoon, 2013). Furthermore, compared with quantitative *meta*-analysis, which uses additive logic to synthesize findings, qualitative *meta*-analysis interprets key qualitative evidence from different contexts to achieve higher replicability of theories (Edmondson & McManus, 2007).

The research design of qualitative *meta*-analysis is particularly suitable for our study for the following reasons. Firstly, a crucial insight from our literature review is that most research on ecosystem orchestration remains qualitative in nature, and case studies make up a large proportion of the qualitative research. Secondly, ecosystem orchestration is complex and ambiguous. Although a range of case studies have explored ecosystem orchestration, providing rich, contextualized empirical descriptions, the existing results remain relatively isolated and lack systematicity. Instead, qualitative *meta*-analysis facilitates the provision of general inferences, achieving a comprehensive understanding that transcends the results of individual studies (Rauch et al., 2014). To our knowledge, there is no study in the domain of ecosystem orchestration that has tried to employ this method. In conclusion, qualitative *meta*-analysis is most beneficial in moderate or mature research areas with unique phenomena and lack of adequate quantitative measures (Edmondson & McManus, 2007), which is exactly in line with the current state of research in the field of ecosystem orchestration.

Inspired by previous research in the domain of qualitative *meta*-analysis (Hoon, 2013; Laubengaier et al., 2022) and by modifying, integrating, and localizing some of the phases according to the conditions of our study, we defined a four-stage process to conduct a qualitative meta- analysis. The initial stages encompass a thorough *search in databases* and *manual filtering* in the data source section, followed by careful *within-case analysis* and *cross-case analysis* in the data analysis section. Each stage will be explained in detail.

*3.2. Data sources*

The first and pivotal step is to identify literature that is closely related to our research question. As depicted in Fig. 1, we conducted a comprehensive search in databases to locate relevant research materials. Subsequently, we established rigorous inclusion and exclusion criteria during the manual filtering stage. The literature was filtered progressively by subject area, document type, document language, research method, and research theme.

*3.2.1. Stage one: Search in databases*

This study undertook a rigorous literature retrieval process to gather relevant articles on ecosystem orchestration from both Web of Science (WoS) Core Collection and Scopus databases. To ensure high-quality papers, we focused on editions of Science Citation Index Expanded and Social Science Citation Index within WoS. The search term "*ecosystem orchestrat\**" was utilized, covering variations of "*ecosystem orchestration*", "*ecosystem + orchestrator*" and "*ecosystem + orchestrate*". The search was limited to publications up to December 2022. Initially, 371 publications were retrieved from WoS, and Scopus yielded 850 publications. The subject area was then narrowed down to business, management, and accounting, resulting in 112 papers retained from WoS and 202 from Scopus. We filtered the document types to include only articles, review articles, and conference papers, eliminating book chapters and books. This step left us with 178 papers from Scopus, while no exclusions were made in WoS. Furthermore, we only included publications written in English, and after excluding one paper in French, a total of 177 papers from Scopus were retained. Next, we aggregated the 289 papers from both databases and removed 96 duplicates, resulting in a final selection of 193 unique papers. These 193 documents form the foundation for the subsequent manual filtering stage.

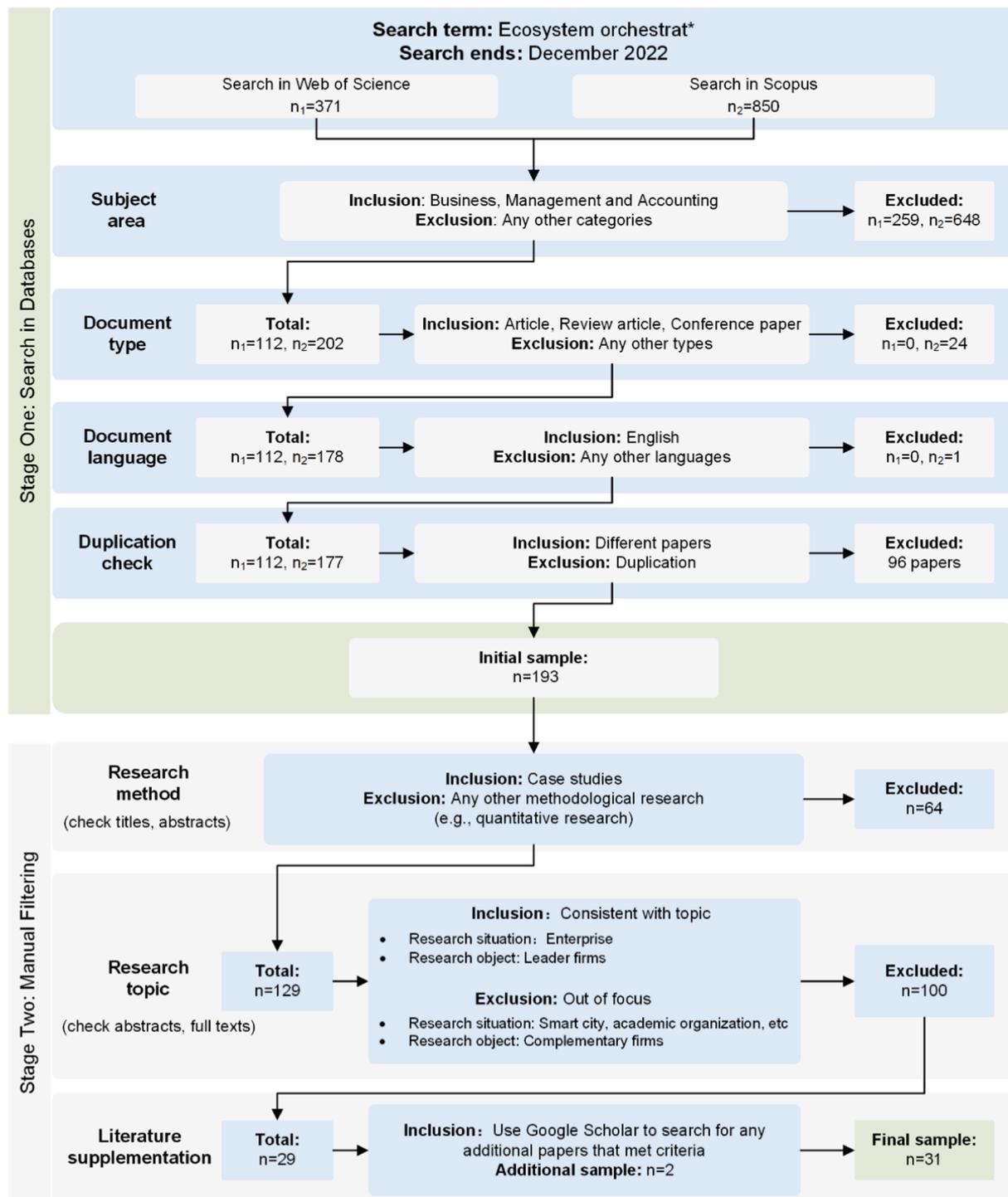

**Fig. 1.** Sampling procedure.

*3.2.2. Stage two: Manual filtering*

The manual filtering stage consists of three main parts: screening based on research method, screening based on research topic, and literature supplementation. Firstly, the focus of the research method filtering was on identifying papers that utilized a case study methodology, while excluding quantitative studies, conceptual studies, and literature reviews that did not meet the specified research method criteria. Initially, 115 articles were found to satisfy the standards based on their titles and abstracts, while 28 articles required further evaluation through reading their full texts. After this comprehensive assessment, 14 articles were deemed to meet the criteria, resulting in 129 qualitative case studies being retained. Secondly, in the filtering process for research topic, we applied specific inclusion and exclusion criteria to ensure that the selected papers aligned with the requirements of the research theme. Specifically, we focused on cases within enterprise scenarios, excluding cases in smart cities or academic organizations. Additionally, we retained essays that focused on ecosystem orchestrators, while excluding papers from other perspectives, such as complementary companies. Importantly, the selected papers needed to address our study's research question. These screenings yielded a set of 29 papers that met all the established requirements.

Thirdly, to ensure the completeness of our selection, we further utilized Google Scholar to search for any additional papers that met our established criteria. This supplementary search contributed to the discovery of two additional papers that satisfied our criteria. Consequently, the final selection of papers for our study on ecosystem orchestration comprised a total of 31 published case studies (see Table 1) and cover 111 business cases. This is a reasonable number of studies for in-depth analysis and interpretation according to Laubengaier et al. (2022). It also satisfies the optimal data range requirements (12–100) for conducting qualitative *meta*-analysis, as proposed by Timulak (2009). Moreover, the number of studies is in close proximity to those conducted by Berente et al. (2019) (35 studies), Laubengaier et al. (2022) (35 studies), Küberling-Jost (2021) (20 studies), and Habersang et al. (2019) (43 studies).

In conclusion, we conducted a comprehensive search in the WoS and Scopus databases, supplemented the results with Google Scholar, and carefully evaluated the selected papers by reading the titles, abstracts, and full texts to ensure the inclusion of relevant papers that met our predefined requirements.

### 3.3. Data analysis

The data analysis in this study draws from the research of Hoon (2013) and Laubengaier et al. (2022), and it encompasses both within-case and cross-case analysis. Within-case analysis involves conducting individual content analyses for each case to identify first-order codes. These first-order codes are then synthesized during the cross-case analysis to form second-order themes, and these themes are further aggregated into dimensions. Inspired by the research of Habersang et al. (2019) and Küberling-Jost (2021), our data analysis used abductive reasoning, which combines the features of deductive and inductive reasoning and facilitates scholars to build theories in the qualitative secondary analysis (Vila-Henninger et al., 2022). It is important to note that qualitative *meta*-analysis does not rely on reusing the first-hand data collected by the original researchers. Instead, it builds upon their interpretations and understanding of the data (Hoon, 2013). Moreover, we specifically focus on the findings, discussion, and conclusion sections of previous case studies to explore novel theoretical insights.

#### 3.3.1. Within-case analysis

In the within-case analysis, we used both deductive and inductive coding to identify first-order codes. We start by developing a deductive coding scheme based on the existing theoretical framework of ecosystem orchestration. Resource-based view (Barney, 1991), dynamic capability (Teece et al., 1997) and collective action theory (Olson, 1965) are generally regarded as important theoretical foundations in the field of ecosystem orchestration, so we chose them as the theoretical background. To create a comprehensive deductive coding system, we then analyzed pertinent literature about these three theories, which contributed to identify relevant deductive codes (see Table 2). Subsequently, the deductive coding scheme was applied as a guidebook to further perform deductive coding in 31 papers extracted form data source. By using the deductive coding system, we systematically filtered the data to find first-order codes that can answer our research question. At the same time, inductive coding or open coding (Gioia et al., 2013) was conducted to explore additional information that was not preconceived, resulting in the generation of inductive codes. In this coding process, we focus on the researchers' descriptions in the original studies about the specific practices or activities of orchestrators in orchestrating ecosystem. Besides, we also paid attention to variables that affect ecosystem orchestration and those that are affected by ecosystem orchestration, as well as those that influence each other, appear together, or appear sequentially (Küberling-Jost, 2021; Laubengaier et al., 2022). This stage generated a total of 377 first-order concepts,

**Table 1**
Case studies in this paper.

| No. | Paper | Research question | Industry/firms |
|---|---|---|---|
| 1 | Zeng et al. (2022) | How do platform-based entrepreneurial companies orchestrate resources to achieve scalability within ecosystems? | Platform-based entrepreneurial firms (1 case) |
| 2 | Sjödin et al. (2022) | How do equipment suppliers tailor ecosystem strategies to achieve digitally enabled process innovation in diverse industrial customer settings? | Manufacturers (8 cases) |
| 3 | Poblete et al. (2022) | How do various temporal logics impact the capacity of keystone actors to efficiently orchestrate an innovation ecosystem? | Construction industry (1 case) |
| 4 | Mann et al. (2022) | How does a focal firm assume the role of orchestrator in driving digital transformation within its business ecosystem? | Security services (1 case) |
| 5 | Lingens, Seeholzer, et al. (2022) | How do orchestrators utilize various configurations of complementarities within emerging ecosystems | Moving, home, sharing, car, 3d-printed sole, insurance, mobile payment, factoring (8 cases) |
| 6 | Franco et al. (2022) | What are the primary capabilities that a luxury hotel must cultivate when operating as the orchestrator of a local gastronomic business ecosystem? | Hotel (1 case) |
| 7 | Cui & Han (2022) | How does a focal enterprise execute resource orchestration in its ecosystem strategy to ensure sustainability in the digital era? | Video-streaming industry (1 case) |
| 8 | Cui et al. (2022) | How do keystone actors govern their business ecosystems when faced with conditions of both resource sufficiency and resource insufficiency? | Internet brand ecological operation group (1 case) |
| 9 | Blackburn et al. (2022) | What are the orchestration mechanisms employed by platform orchestrators in *meta*-organizations aiming for circular value creation? | Agrifood, manufacturing, services (10 cases) |
| 10 | Aagaard & Rezac (2022) | How can orchestrators govern the interplay of inter-organizational relationship mechanisms in open innovation projects across ecosystems? | Technology company (1 case) |
| 11 | Zeng et al. (2021) | How do sharing economy platforms manage their resources to create value and build competitive advantages? | Transportation, retail, and food sectors (6 cases) |
| 12 | Tian et al. (2021) | What are the fundamental building components that facilitate the co-evolution and co-creation dynamics within a platform, and how do these components enable platform development and utilization? | Textile industry (4 cases) |
| 13 | Marheine et al. (2021) | How does an incumbent telecoms operator shift its role from being an IoT enabler to an orchestrator in the ecosystem? | Telecoms industry (1 case) |

**Table 1** (*continued*)

| No. | Paper | Research question | Industry/firms |
|---|---|---|---|
| 14 | Lingens, Miéhé et al. (2021) | How does the design of an ecosystem take shape based on the surrounding conditions? | Digital services (10 cases) |
| 15 | Lingens & Huber (2021) | How do firms assign orchestrator tasks to particular participants within the ecosystem? | Media, insurance, mobility, security, finance, logistics, public transportation, home (10 cases) |
| 16 | Lingens, Böger et al. (2021) | How can a startup undertake the tasks of an orchestrator and overcome challenges? | Finance, insurance, sports equipment, home, mobility, analytics, energy, security (9 cases) |
| 17 | Kamalaldin et al. (2021) | How do equipment suppliers configure appropriate ecosystem strategies to achieve digitally enabled process innovation across different industrial customer contexts? | Equipment suppliers (6 cases) |
| 18 | Gomes et al. (2021) | How do companies manage dispersed knowledge in ecosystems? | Car part supplier, ride sharing company, energy transmission company, hospital, start-up of health sector, digital firm connecting gyms (6 cases) |
| 19 | Cui et al. (2021) | How does a traditional intermediary transition into an e-intermediary by leveraging strategic logic and resource orchestration? | An e-commerce platform (1 case) |
| 20 | Reypens et al. (2021) | How do orchestrators mobilize network members in multi-stakeholder networks across organizational boundaries? | Medical (1 case) |
| 21 | Masucci et al. (2020) | How can firms strategically orchestrate outbound open innovation to accelerate technological advancements among their collaborating partners? | Oil and gas industry (5 cases) |
| 22 | Zucchella & Previtali (2019) | How to get a better understanding of a business model built on circular principles? | Waste recycling and upcycling in agriculture (1 case) |
| 23 | Parida et al. (2019) | How do manufacturing companies orchestrate ecosystem-wide transformation towards the circular economy paradigm? | Manufacturing companies (6 cases) |
| 24 | Laczko et al. (2019) | How does the central actor enhance the sustainability of a multi-stakeholder platform within the sharing economy? | B2B online sharing platform (1 case) |
| 25 | Leten et al. (2013) | What central role does the IP model play in the orchestrator's efforts to manage and expand an innovation ecosystem? | A public research institute in nano-electronics (1 case) |
| 26 | Still et al. (2014) | How can data-driven network visualizations be employed to generate insights for orchestrating an innovation ecosystem? | ICT Labs (1 case) |
| 27 | Tabas et al. (2022) | What various types of orchestrator roles can be identified within the entrepreneurial health tech ecosystem, and what bundles of role-specific capabilities are associated with them? | Health technology (1 case) |
| 28 | Lu & Zhang (2022) | How do ecosystem actors orchestrate ecosystem- | Seaweed industry (1 case) |

**Table 1** (*continued*)

| No. | Paper | Research question | Industry/firms |
|---|---|---|---|
| | | specific resources and capabilities to facilitate the growth of new ventures in the context of Hub-based entrepreneurial ecosystems? | |
| 29 | Azzam et al. (2017) | How do focal firms impact business ecosystem stability through patent management | Aerospace industry (1 case) |
| 30 | Ritala et al. (2013) | What actions can the orchestrator undertake, and what mechanisms and structures can be employed to ensure value creation and capture among the participants in their innovation ecosystems? | Electronics and information and communication technology & aerospace and defense sectors (2 cases) |
| 31 | Isckia et al. (2020) | How to orchestrate platform ecosystems to ensure the commercialization of ongoing innovation flows? | Technology (3 cases) |

including 24 deductive codes and 353 inductive codes.

*3.3.2. Cross-case analysis*

In the cross-case analysis, we conducted axial coding to group the above first-order concepts into second-order themes and then classify dominant themes into different dimensions. Through an iterative process of comparing and contrasting codes for similarities and differences, we extracted, condensed, and summarized these themes, ultimately identifying dominant second-order patterns. Thereafter, we explored associations between themes and identified logical relationships among them, facilitating the aggregation of interrelated themes into dimensions. The whole data analysis process was iterated over and over again to find meaningful and logical results. Finally, we identified 5 dimensions, comprising a total of 15 s-order themes and 30 first-order codes.

To effectively present our findings, we created a data structure (see Fig. 2) that illustrates the specific contents and affiliations of first-order codes (sub-orchestration activities), second-order themes (key orchestration activities), and aggregate dimensions (ecosystem orchestration practices). Subsequently, we identified the relationships between these aggregate dimensions and developed theoretical frameworks and arguments to address our research questions (Gomes, dos Santos, et al., 2022). Based on these discussions, we propose a complete framework for ecosystem orchestration practices (see Fig. 3).

## 4. Findings

In the following section, we present the findings related to our research question. Table 3 provides an overview of the findings and their illustrative sources. After synthesizing the existing fragmented research, we have identified five ecosystem orchestration practices that are carried out by the ecosystem orchestrator based on our data analysis. These practices include ecosystem strategic design practices, ecosystem relational practices, ecosystem resource integration practices, ecosystem technological leveraging practices, and ecosystem innovation practices, which are further detailed in Fig. 2.

### 4.1. Ecosystem strategic design practices

Ecosystem strategic design practices serve as both a roadmap and a strategic blueprint to guide the future development and growth of the ecosystem. Our data analysis underscores the pivotal role these practices play in ecosystem orchestration. Ecosystem strategic design is dynamic, with its optimal configuration varying according to the prevailing environmental uncertainty (Lingens, Miéhé, et al., 2021). Orchestrators proactively engage in designing the value proposition, forming an ecosystem identity, and monitoring growth and changes.

**Table 2**
Deductive coding scheme.

| Theoretical background | Central themes | 1st order concepts (deductive codes) | Exemplary Papers |
| --- | --- | --- | --- |
| *resource-based view* (orchestrators use valuable, rare, inimitable, and non-substitutable resource to exploit opportunities, neutralize threats and gain sustained competitive advantages) | physical capital resources; human capital resources; organizational capital resources (bundles of tangible and intangible assets) | conceptualize and implement strategies using available resources; acquire distribution channels ; build good relationships with customers ; establish a solid reputation; gain insight into strategic opportunities; develop strategic plans for identifying and leveraging resources | Barney (1991); Barney et al. (2001); Barney (2001) |
| *dynamic capability* (orchestrators develop the capabilities of sensing, seizing, and reconfiguring to adapt effectively and efficiently, and renew internal and external firm competencies and resources in a rapidly changing environment) | sensing opportunities and threats; seizing opportunities; managing threats and reconfiguration | promote the emergence of a common vision; encourage others to invest in ecosystem-specific initiatives; participating in ad hoc resolution to sustain stability; respond to change as soon as possible; rapid and agile product innovation; coordination and redeployment of internal and external competence; identification of difficult-to-imitate competences; integration of external activities and technologies; coordination of organizational routines; facilitation of inter-organizational learning; sense needs and complete reconfiguration and transformation continuous scan and evaluate the market and technology; decentralization and local autonomy | Teece et al. (1997); Teece (2007); Teece (2012); de Miguel et al. (2022); Ogunrinde (2022); Foss et al. (2023) |
| *collective action theory* (orchestrators, internal participants, and external actors collectively drive ecosystem development) | interaction between various participants; collective action; achieve common goals | deliver outputs collectively that exceed what any single participant could achieve independently; develop a collective identity; interact with others to formulate a shared goal; increase reputations and expected reciprocal actions; motivate cooperation by advancing personal | Olson (1965); Hargrave & Van de Ven (2006); Laamanen & Skålén (2014); Zhao et al. (2014); Thomas & Ritala (2022) |

**Table 2** (*continued*)

| Theoretical background | Central themes | 1st order concepts (deductive codes) | Exemplary Papers |
| --- | --- | --- | --- |
| | | profit; share practices to value co-create | |

The first key orchestration activity, ***designing the value proposition***, involves defining the ecosystem's objectives, blueprint, and unique position vis-à-vis other ecosystems (Panico & Cennamo, 2022). The value proposition – a central concept in the ecosystem literature – represents the ecosystem's connectivity pattern and establishes its boundaries (Adner, 2017; Linde et al., 2021). Orchestrators must make strategic decisions in designing an appealing value proposition (Isckia et al., 2020), playing a crucial role in defining a shared vision and translating it into a business model (Zucchella & Previtali, 2019). On the one hand, orchestrators *negotiate the value offering* with partners, giving due consideration to participants' motivations and ensuring fair value distribution. This requires finding, setting, and maintaining the right balance between the common ecosystem vision and partners' self-interest to incentivize active participation and contribution (e.g., double value proposition) (Isckia et al., 2020; Kamalaldin et al., 2021; Laczko et al., 2019). Joint forums and meetings are effective in achieving this balance (Reypens et al., 2021; Ritala et al., 2013). This negotiation and interaction process not only safeguards ecosystem stability but also aligns heterogeneous complementors with the joint value proposition (Aagaard & Rezac, 2022; Mann et al., 2022), which is beneficial for seizing and exploiting business opportunities (Linde et al., 2021). On the other hand, orchestrators and partners *share a common vision of shared value*, which is central to decision making and is aimed at reaping long-term benefits (Ritala et al., 2013). Achieving this common vision facilitates communication, builds trust, and strengthens commitment and engagement among partners (Zucchella & Previtali, 2019), ensuring a stable and sustainable relationship. Significantly, it shifts the mindset of actors from a traditional value chain towards complementary co-creation (Aagaard & Rezac, 2022; Jacobides et al., 2018).

In addition to designing the value proposition, our research has uncovered another crucial facet of ecosystem orchestration: ***forming an ecosystem identity***. This is defined as the collective understanding among participants on the fundamental, enduring, and distinctive attributes that define an ecosystem's value proposition (Thomas & Ritala, 2022). It encompasses elements such as distinctive features, branding, and cultural attributes, which serve to define the specific traits of the ecosystem. Orchestrators strive to *create these unique identity characteristics*, making strategic choices to differentiate the ecosystem from competitors and attract aligned participants (Blackburn et al., 2022). However, building a credible identity can be challenging, often requiring assistance from partners who bring in new perspectives and expertise (Tabas et al., 2022). This identity formation process involves *legitimizing the identity* – in other words, winning the agreement of others. Ecosystem legitimacy emerges as a dynamic process where orchestrators, participants, and external actors engage in collective action (Thomas & Ritala, 2022). It fosters trust and a sense of belonging within the ecosystem, enhancing internal and external understanding and recognition. Ecosystem identity is intrinsically intertwined with the identities of individual participants. In emerging markets, ecosystem identity building may leverage the brand of an established ecosystem pilot participant as a foundational element (Blackburn et al., 2022).

Finally, the third key orchestration activity is ***monitoring growth and changes***, emphasizing the detection of and adaptation to external dynamic changes. To effectively orchestrate an ecosystem, orchestrators must *scan and evaluate market and technological conditions*, identifying emerging opportunities, mitigating risks, and responding swiftly to changes. For instance, orchestrators continuously gather and iterate consumer feedback to meet their needs (Mann et al., 2022). These opportunities are strategically leveraged by orchestrators to drive growth and foster innovation. The orchestrator's role in ecosystem development lies in the ability to continuously explore, assess, and act on emerging opportunities to create and deliver value (Laczko et al., 2019). Additionally, *orchestrators gain strategic foresight on emerging issues*, which they use to direct actions towards orchestrating resource pools and spur rapid growth (Zeng et al., 2022). These monitoring activities involve not only a general assessment of the business landscape but also careful scrutiny of potential opportunities for targeted initiatives concerning specific objectives and customers (Kamalaldin et al., 2021).

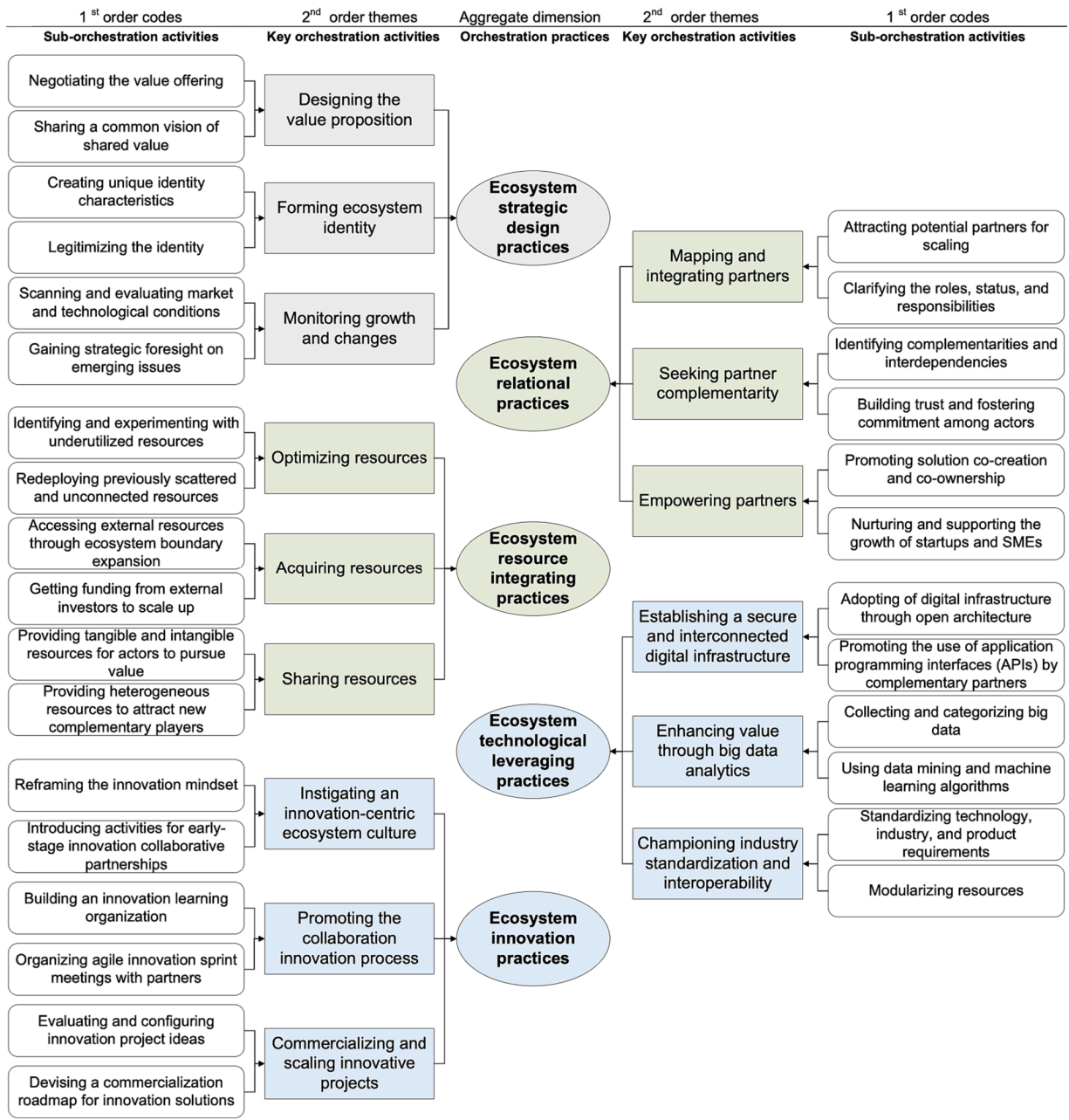

**Fig. 2.** Data structure-ecosystem orchestration practices.

The above-mentioned discussion leads to our first proposition:

**Proposition 1.** *In orchestrating ecosystems, orchestrators initiate comprehensive ecosystem strategic design practices, including designing the value proposition, forming an ecosystem identity, and monitoring growth and changes. These activities act as facilitators of value creation, identifying the specific value to be created and directing other orchestrated actions accordingly.*

*4.2. Ecosystem relational practices*

Our study reveals that ecosystem orchestrators engage in what we label as ecosystem relational practices, comprising three vital factors that are detailed subsequently. Ecosystems consist of relationships that transcend collections of bilateral interactions (Adner, 2017). Ecosystem partners, acting as complementors, provide crucial assistance to orchestrators in securing financial, human, and technical resources, bridging technological gaps, and adapting to rapid changes (Kolagar et al., 2022). When partners share aligned interests and promote altruistic behavior, value creation becomes a joint responsibility of the entire network, not solely dependent on orchestrators' capabilities and resources (Laczko et al., 2019). Hence, managing relationships is a pivotal aspect of ecosystem orchestration.

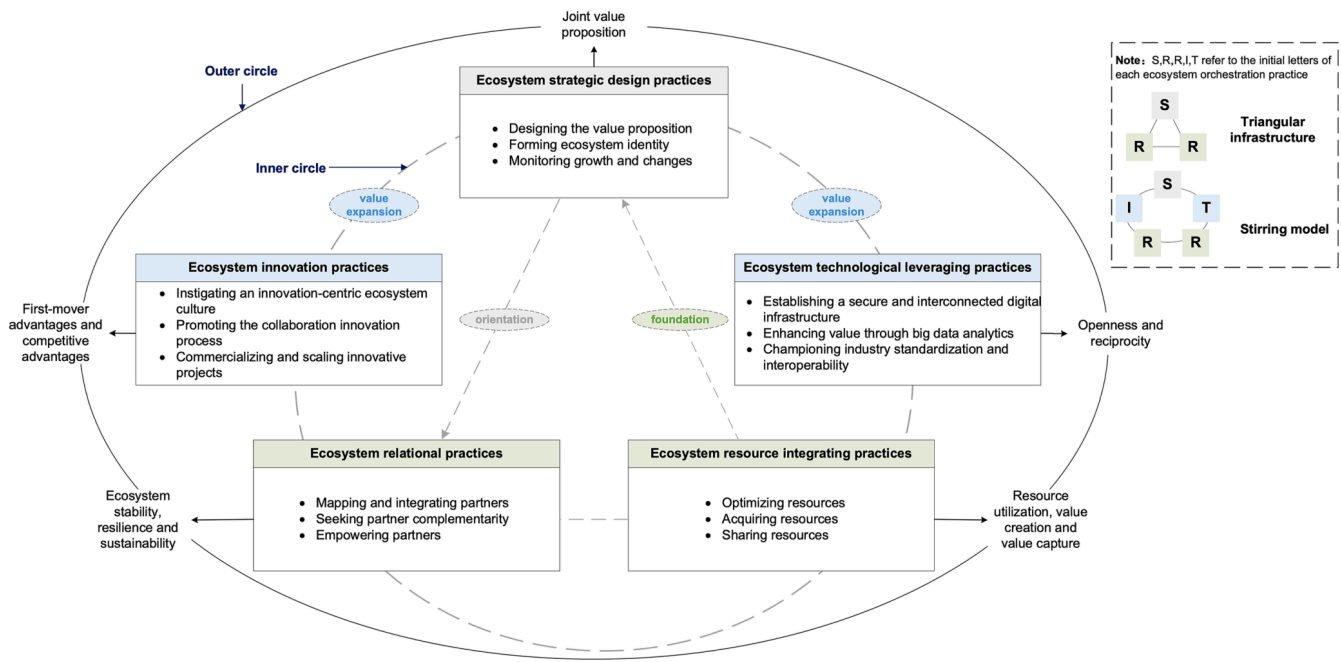

**Fig. 3.** Stirring model of ecosystem orchestration practices.

The first key activity, ***mapping and integrating partners***, involves identifying partners and creating a network that aligns with the ecosystem's value proposition. External companies, particularly from other industries, add diversity and creativity to the ecosystem (Lu & Zhang, 2022), enhancing its existing value proposition and facilitating entry into emerging businesses (Lu & Zhang, 2022; Miehé et al., 2022). Orchestrators use various means to *attract potential partners for scaling*, such as leveraging patents and other intellectual property assets to draw in new actors, given their inherently proprietary nature (Azzam et al., 2017). As the ecosystem expands, controlling the entry of new partners based on risk and benefit analysis becomes crucial (Parida et al., 2019). Orchestrators *clarify the roles, status, and responsibilities* of internal partners, making their place in the ecosystem explicit (Aagaard & Rezac, 2022).

Drawing on initial insights and a deeper understanding of participants, ecosystem orchestrators strive to ***seek partner complementarity*** to align and secure partners for long-lasting collaboration. Trust and the maturity of symbiotic relationships are crucial for thriving ecosystems (Ritala et al., 2013). In newly established ecosystems where participating companies are relatively unfamiliar with each other, orchestrators assume the role of matchmakers, facilitating the harmonious development of partnerships. They *identify complementarities and interdependencies* between actors for subsequent collaboration (Zucchella & Previtali, 2019). These efforts optimize the ecosystem's partner layout by selecting, positioning, and connecting partners, thereby leveraging the complementarities and network effects in the case of platform ecosystems (Aagaard & Rezac, 2022; Jovanovic, Sjödin, et al., 2022; Zeng et al., 2022). Building on this initial recognition of actor complementarities and interdependencies, orchestrators can enhance relational interdependence through structural complementarity (Lingens, Seeholzer, et al., 2022). Another strategy to promote interdependence focuses on emotional connection, especially in *building trust and fostering commitment among actors* (Tian et al., 2021). Long-term relationships are more beneficial for companies (Omer, 2022), making the establishment of regular routines, such as face-to-face meetings, vital for enhancing communication and strengthening partner bonds (Reypens et al., 2021). These activities encourage and facilitate cooperation among previously isolated actors in the ecosystem, thereby retaining partners for its stable and sustainable development. Thus, partners can effectively combine and utilize their distinct resources in a complementary fashion (Linde et al., 2021).

As the number of heterogeneous actors, transactions, and interactions increases, so does the risk of misalignment in the outcomes desired (Carida et al., 2022). Consequently, uniting partners around a common goal from the top down becomes increasingly challenging (Reypens et al., 2021). Eventually, the complexity of the situation may surpass the orchestrator's capacity to manage it alone (Isckia et al., 2020; Kolagar et al., 2022). This scenario introduces another crucial function of orchestrators, termed ***empowering partners***. This function underscores the importance of providing support and inspiration to partners to co-create value. Orchestrators make significant upfront investments to alleviate partner uncertainty (Parida et al., 2019) and enhance their risk resilience. Additionally, orchestrators deploy financial resources to motivate participants, establishing direct financial incentives and indirectly spurring motivation by investing in technology and other appealing resources (Gomes et al., 2021; Masucci et al., 2020). It is important to note that ecosystem partners are not merely beneficiaries but also value creators within the ecosystem. Successful ecosystem orchestration is attainable only when a company actively involves its partners in the process (Sjödin et al., 2022). Ecosystem orchestrators *promote solution co-creation and co-ownership* (Sjödin et al., 2022) and even encourage other actors to participate in co-orchestration (Reypens et al., 2021). Mann et al. (2022) observed that orchestrators empower all business ecosystem actors to leverage opportunities for developing and testing innovative digital applications. Special attention is required for small and medium-sized enterprises (SMEs), a critical category of actors. Orchestrators play a pivotal role in *nurturing and supporting the growth of startups and SMEs* by endowing these firms with credibility and legitimacy (Tabas et al., 2022). Moreover, they provide essential support to SMEs in their efforts to enter new markets (Lingens, Seeholzer, et al., 2022). These nurturing actions are instrumental in fostering the development of new business opportunities within the ecosystem (Parida et al., 2019).

Thus, we propose:

**Proposition 2.** *In the orchestration of ecosystems, orchestrators are required to undertake ecosystem relational practices, including mapping, engaging, and integrating partners, seeking partner complementarity, and empowering partners. These activities form the basis of ecosystem*

**Table 3**
Overview of findings and illustrative sources.

| Key findings | | Illustrative sources |
|---|---|---|
| Ecosystem strategic design practices | Designing the value proposition | Aagaard & Rezac (2022), Isckia et al. (2020), Kamalaldin et al. (2021), Laczko et al. (2019), Mann et al. (2022), Reypens et al. (2021), Ritala et al. (2013), and Zucchella & Previtali (2019) |
| | Forming an ecosystem identity | Blackburn et al. (2022) and Tian et al. (2021) |
| | Monitoring growth and changes | Kamalaldin et al. (2021), Laczko et al. (2019), Mann et al. (2022), and Zeng et al. (2022) |
| Ecosystem relational practices | Mapping and integrating partners | Aagaard & Rezac (2022), Azzam et al. (2017), Lu & Zhang (2022), Miehé et al. (2022), and Parida et al. (2019) |
| | Seeking partner complementarity | Aagaard & Rezac (2022), Linde et al. (2021), Lingens, Seeholzer, et al. (2022), Reypens et al. (2021), Ritala et al. (2013), Tian et al. (2021), Zeng et al. (2022), and Zucchella & Previtali (2019) |
| | Empowering partners | Gomes et al. (2021), Lingens, Seeholzer, et al. (2022), Mann et al. (2022), Masucci et al. (2020), Parida et al. (2019), Reypens et al. (2021), Sjödin et al. (2022), and Tabas et al. (2022) |
| Ecosystem resource integrating practices | Optimizing resources | Blackburn et al. (2022), Cui et al. (2022), Mann et al. (2022), Zeng et al. (2021), and Zeng et al. (2022) |
| | Acquiring resources | Cui et al. (2022), Cui & Han (2022), Lingens, Böger, et al. (2021), Poblete et al. (2022), and Zeng et al. (2022) |
| | Sharing resources | Azzam et al. (2017), Cui et al. (2022), Cui et al. (2021), Isckia et al. (2020), Kamalaldin et al. (2021), Leten et al. (2013), Lu & Zhang (2022), and Tabas et al. (2022) |
| Ecosystem technological leveraging practices | Establishing a secure and interconnected digital infrastructure | Aagaard & Rezac (2022), Blackburn et al. (2022), Isckia et al. (2020), Kamalaldin et al. (2021), Lingens, Böger, et al. (2021), Marheine et al. (2021), and Zeng et al. (2022) |
| | Enhancing value through big data analytics | Kamalaldin et al. (2021), Still et al. (2014), Zeng et al. (2021), and Zeng et al. (2022) |
| | Championing industry standardization and interoperability | Azzam et al. (2017), Gomes et al. (2021), Lingens, Böger, et al. (2021), Lu & Zhang (2022), Parida et al. (2019), Schreieck et al. (2021) |
| Ecosystem innovation practices | Instigating an innovation-centric ecosystem culture | Aagaard & Rezac (2022), Lingens & Huber (2021), Lingens, Miehé, et al. (2021), Lingens, Seeholzer, et al. (2022), Lu & Zhang (2022), Mann et al. (2022), Reypens et al. (2021), Sjödin et al. (2022), Zeng et al. (2021), |

**Table 3** (*continued*)

| Key findings | | Illustrative sources |
|---|---|---|
| | | and Zucchella & Previtali (2019) |
| | Promoting the collaboration innovation process | Laczko et al. (2019), Leten et al. (2013), Lingens & Huber (2021), Masucci et al. (2020), Poblete et al. (2022), Ritala et al. (2013), Sjödin et al. (2022), Tian et al. (2021), and Zucchella & Previtali (2019) |
| | Commercializing and scaling innovative projects | Aagaard & Rezac (2022), Gomes et al. (2021), Isckia et al. (2020), Poblete et al. (2022), and Reypens et al. (2021) |

*orchestration and contribute to ecosystem stability, resilience, and sustainability.*

*4.3. Ecosystem resource integrating practices*

Numerous enterprises strategically opt to join an ecosystem primarily to alleviate their resource scarcity. This participation facilitates access to crucial resources and reaps the benefits of prospective economic advantages associated with resource efficiency (Blackburn et al., 2022; Kolagar et al., 2022). Ecosystem orchestrators, positioned as central entities in the ecosystem, are often perceived as resource abundant, possessing control over the most indispensable resources (Cui et al., 2019). Given the diversity of resources and entangled partner resources, it is crucial for orchestrators to dynamically and effectively reconfigure these resource bundles (Isckia et al., 2020). The third identified practice, termed ecosystem resource integrating practices, includes optimizing existing resources, acquiring new external resources, and facilitating the mutual use of resources by partners.

***Optimizing resources*** emphasizes the identification, reorganization, and alignment of existing resources to co-create value (Carida et al., 2022). Based on their acquaintance with partners, orchestrators strategically match resources to necessary tasks (Reypens et al., 2021). On the one hand, they *identify and experiment with underutilized resources*, such as idle assets, to retain resource value (Zeng et al., 2022). Particularly in situations with limited resources, orchestrators should focus on internal idle resources, redeploying them flexibly within their ecosystem to address resource dilemmas and foster flexibility capabilities (Cui et al., 2022). On the other hand, orchestrators *redeploy previously scattered and unconnected resources* to expand their resource portfolio, seek breakthroughs, and change the market (Zeng et al., 2021). For instance, orchestrators merge the dispersed resources of partners to enhance resource usage efficiency (Mann et al., 2022). Additionally, adopting a circular resource strategy, encompassing principles of reuse, reduction, recycling, and recovery, is highly effective (Blackburn et al., 2022). These activities demonstrate that resource optimization is a continuous, dynamic process that perpetually seeks optimal value creation.

The paradigm of resource orchestration is undergoing a significant transformation, shifting from traditional management of static internal resources to a more holistic, dynamic, and evolving approach that encompasses both internal and external resources (Zeng et al., 2021). ***Acquiring resources*** involves efforts by ecosystem orchestrators to assist partners in accessing complementary external resources. Orchestrators identify any misalignments between existing and required resources and subsequently *access external resources through ecosystem boundary expansion* to meet complementors' needs (Cui & Han, 2022). They can purchase resources directly from strategic factor markets or internalize external resources through cooperation. In addition, ecosystem orchestrators can obtain funding from external investors (Lingens, Böger, et al., 2021). Lingens, Böger, et al. (2021) note that, when startups assume the role of an orchestrator, they often require *funding from external investors to scale up*. Furthermore, continuously updating resources in the ecosystem is essential to maintain competitiveness and reduce vulnerability (Zeng et al., 2021). As Zeng et al. (2022) explain, the strategic emphasis of these resource-acquiring processes has shifted from possessing and controlling all valuable resources to attracting, accessing, cultivating, and mutually exchanging external resources. Indeed, orchestrators should strive to balance temporary external resources with enduring internal resources (Poblete et al., 2022).

**Sharing resources** involves actively promoting openness and the sharing of resources to stimulate innovation and value creation (Isckia et al., 2020; Teng et al., 2023). Knowledge, capital, and other critical factors collaborate in a system to create a symbiotic and continuous environment (Hu & Zhang, 2023). Some resources, previously invaluable to one organization, may be useful to others (Burström et al., 2021). From the perspective of partners, their motivation to share resources varies. One reason is to access a larger pool of resources by collaborating with other ecosystem actors, potentially leading to a "piece of the bigger pie" (Kamalaldin et al., 2021). Another reason is to safeguard essential resources amidst internal and external uncertainties, which may necessitate selective or limited collaboration with other players in the ecosystem (Kamalaldin et al., 2021). In ecosystems, orchestrators are responsible for facilitating resource flow, *providing tangible and intangible resources for actors to pursue value* (Lu & Zhang, 2022). For instance, Leten et al. (2013) assert that ecosystem orchestrators consistently invest in intellectual property within their field of expertise and share it with partners, stimulating ecosystem progress. Moreover, ecosystem orchestrators can offer specific investments to support weaker players in the ecosystem (Tabas et al., 2022). Ecosystem innovation is triggered by sharing and combining diverse intangible and tangible resources among organizations (Poblete et al., 2022). Furthermore, *orchestrators provide heterogeneous resources to attract new complementary players* and leverage business opportunities (Cui et al., 2022). The creation and capture of value are enhanced when ecosystem participants' resources are consolidated (Kamalaldin et al., 2021). In these activities, the orchestrator acts as a broker, not aiming to own the resources but to capture a portion of the value generated from transactions and exchanges (Blackburn et al., 2022).

Thus, we propose:

**Proposition 3.** *In orchestrating ecosystems, orchestrators are required to undertake ecosystem resource integrating practices, including optimizing resources, acquiring resources, and sharing resources. These activities comprise the basis of ecosystem orchestration and contribute to ensuring efficient resource utilization, value creation, and value capture.*

*4.4. Ecosystem technological leveraging practices*

In the context of a flourishing technological revolution, digital technology has become the primary arena for innovation, bringing about significant developments in business (Chen et al., 2022; Dana et al., 2022; Kanski & Pizon, 2023; Shang et al., 2023; Tu et al., 2023). With the rapid advancement of digital technologies, such as big data analytics, the Internet of things, artificial intelligence, blockchain, and machine learning, digitization has opened up new opportunities for ecosystems (Jovanovic, Sjödin, et al., 2022). These opportunities facilitate exchanging information, building competitiveness, and innovating business models (Cui et al., 2021; Kamalaldin et al., 2021; Miehé et al., 2022). There is a noticeable trend among traditional manufacturers to actively seek strategic ecosystem collaboration with digital technology and software providers to create digital value (Sjödin et al., 2022). These technologies are diffused among organizations and are developed through inter-organizational cooperation (Burström et al., 2021; Fredström et al., 2021; Tagscherer & Carbon, 2023; Xu et al., 2022). For example, the deployment of AI is idiosyncratic, requiring collaborative efforts from various organizational entities to produce difficult-to-imitate AI applications (Mikalef et al., 2023). However, the fragmentation caused by technological advances often results in participants operating in isolated silos, necessitating an orchestrator to efficiently oversee and mediate (Leten et al., 2013). The fourth identified practice is termed ecosystem technological leveraging practices, which refers to orchestrators' utilization of digital technologies to drive the ecosystem.

The primary key activity, termed **establishing a secure and interconnected digital infrastructure**, is guided by a business model that incorporates constant feedback and performance monitoring (Blackburn et al., 2022). Orchestrators foster the *adoption of digital infrastructure through open architecture*, enabling the seamless integration of other ecosystem partners' systems into their digital framework for joint value creation (Kamalaldin et al., 2021). In this digital architecture, other ecosystem actors can connect their digital solutions, identify operational inefficiencies, and enhance customer value (Kamalaldin et al., 2021). Orchestrators can skillfully intertwine technical product and service design with the digital systems, enhancing the reconfiguration potential in the ecosystem (Hsuan et al., 2021; Jovanovic, Sjödin, et al., 2022). Furthermore, the technical architecture and algorithmic management of the ecosystem play a pivotal role in automatically detecting, reporting, and preventing violations of ecosystem rules (Blackburn et al., 2022). It is crucial to recognize that this activity is not a static endpoint but an ongoing and iterative process involving continuous improvement of the technical architecture (Mann et al., 2022). Additionally, orchestrators *promote the use of application programming interfaces (APIs) by complementary partners* (Lingens, Böger, et al., 2021; Zeng et al., 2022). For example, in the e-commerce industry, orchestrators provide partners with access to APIs to enhance their transactional capabilities (Isckia et al., 2020). As highlighted by Marheine et al. (2021), facilitating partner access to devices and developing commercial applications through open and interoperable technologies is beneficial.

Big data is a valuable, rare, irreplaceable, and hard-to-imitate resource, especially in the context of digital trading platforms (Zeng et al., 2021). Ecosystems, with their broad and heterogeneous participant base potentially spanning the entire value chain and product life cycle, provide an opportunity to collect a more extensive set of data. The flow of data between partners in an ecosystem not only deepens the understanding of consumers but also strengthens collaboration in value delivery (Burström et al., 2021). Orchestrators can **enhance value through big data analytics**, which includes the collection and utilization of big data. To initiate this process, *orchestrators collect and categorize big data* on consumer and collaborator behavior based on digital architectures (Still et al., 2014). This data can optimize the data mining algorithms of companies within the ecosystem (Zeng et al., 2022). For instance, cloud-based big data analytic tools enable partners to jointly analyze operational data and proactively identify operational risks (Tian et al., 2021). As data volume reaches a sufficient scale, orchestrators can *use data mining and machine learning algorithms* to transform data into actionable information and stimulate network effects (Kamalaldin et al., 2021; Zeng et al., 2022). Leveraging advanced learning algorithms allows companies to explore a variety of solutions at a lower cost, reducing capital expenditure on designing heavy test samples or developing test sites (Kamalaldin et al., 2021). Furthermore, it aids ecosystem partners in accurately forecasting and meeting consumer behaviors, effectively satisfying market demands and sustainably building competitive advantages (Zeng et al., 2021).

Based on our previous analysis, it is clear that ecosystem partners can access the ecosystem's technological framework by leveraging APIs. However, the inherent technological complexity associated with APIs may present certain limitations in their practical application. Consequently, it is advisable for ecosystem orchestrators to make significant long-term investments in standardizing their technological infrastructure (Aagaard & Rezac, 2022). Orchestrators are committed to **championing industry standardization and interoperability** to reduce access barriers for all ecosystem participants. A pivotal aspect of these orchestration activities involves *standardizing technology, industry, and* *product requirements*, which facilitates cost reduction, operational efficiency, and ecosystem activation (Jovanovic, Kostić, et al., 2022; Lingens, Böger, et al., 2021). Generic technologies are applicable across different markets (Azzam et al., 2017; Cen et al., 2023). Furthermore, Parida et al. (2019) highlight the importance of industrial and technical standards. In this context, ecosystem orchestrators collaborate with key partners to establish informal standards based on mutual commitment, followed by employing formal certification processes to gain broad acceptance among all partners (Parida et al., 2019). Additionally, *modularizing resources* is recognized as a crucial strategy to enhance accessibility and application, thereby lowering the entry barriers that might deter potential partners from embracing digital technologies. To achieve this, technologies must be flexible, allowing reuse in new applications and reconfiguring algorithms and systems to adapt to various user needs and contexts (Sjödin et al., 2021). This modularization involves separating, repackaging, and encapsulating resources, making them easily accessible to ecosystem participants (Lu & Zhang, 2022). Modular resources help ecosystem partners better understand and select the tools and information required to develop applications more efficiently (Schreieck et al., 2021), and these modules can be seamlessly integrated into a broader array of offerings (Sjödin et al., 2021). For example, Gomes et al. (2021) introduce a modularity knowledge strategy where orchestrators can identify different modules and decide which ones to use for co-creation initiatives.

This leads to the following proposition:

**Proposition 4.** *In orchestrating ecosystems, it is recommended that orchestrators adopt ecosystem technological leveraging practices, including establishing a secure and interconnected digital infrastructure, enhance value through big data analytics, and championing industry standardization and interoperability. These activities leverage technological advantages and promote openness and reciprocity within the ecosystem.*

*4.5. Ecosystem innovation practices*

Innovation is a key driver of growth, propelling businesses towards excellence and competitive advantages (Gao et al., 2023; Shahzad et al., 2022). Our analysis reveals that most studies on ecosystem orchestration place significant emphasis on innovation (Carida et al., 2022). Numerous innovations exhibit a systemic nature (Foss et al., 2023), and the concept of generativity is deeply ingrained in ecosystems (Autio, 2022; Thomas & Tee, 2022). Some enterprises actively engage in ecosystems with the expectation of collaboratively creating specific types of innovation (Gomes, dos Santos, et al., 2022), which cannot be achieved by individual enterprises alone. Successful ecosystems are often characterized by collaborative partners who co-develop capabilities centered around pioneering innovations (Sjödin et al., 2022). Ecosystem innovation promotes the formation and development of value proposition-driven ecosystems, ultimately bolstering the competitiveness and sustainability of both individual firms and the entire ecosystem (Linde et al., 2021). The fifth identified practice, termed ecosystem innovation practices, encompasses a set of activities through which ecosystem orchestrators strive to enhance innovation within the ecosystem.

The first key orchestration activity, labeled as **instigating an innovation-centric ecosystem culture**, involves orchestrators *reframing the innovation mindset* within the ecosystem to maintain its innovativeness (Mann et al., 2022). This continuous effort by the orchestrator is crucial for renewing the ecosystem, ensuring its sustainable competitive advantage in a dynamically changing environment. Currently, the perspective on innovation has shifted from firm centric to network centric (Zeng et al., 2021). Orchestrators need to shift partners' mindsets away from traditional value chains towards complementary co-creation, enabled by modularity and a shared vision, fostering an agile mindset that capitalizes on dynamic innovation capabilities (Aagaard & Rezac, 2022) Adopting an agile innovation approach within the ecosystem is essential for ensuring customer centricity and solution resilience (Sjödin et al., 2022). Additionally, there is a need for continual adaptation of existing organizational structures and the redesign of business activities to align with innovative thinking (Lingens, Miehé, et al., 2021). For example, Zucchella and Previtali (2019) suggest that institutional frameworks, business activities, partnerships, and financial systems need to be redesigned to support business model innovation. Considering that exposure to information, knowledge, and opportunities is a prerequisite for innovation (Lingens & Huber, 2021), orchestrators need to *introduce activities for early-stage innovation collaborative partnerships*. This initiative increases the likelihood of partners recognizing and seizing opportunities for innovation. The process involves the careful selection and integration of partners in a joint innovation effort (Lingens, Seeholzer, et al., 2022) and the alignment of various innovation efforts and resources (Reypens et al., 2021). Moreover, orchestrators leverage their expertise to transform pre-established collaborative structures into flexible systems, facilitating the creation of innovative outputs (Reypens et al., 2021). This can be achieved by redesigning and reorganizing processes, regulations, resources, and internal divisions to enable value co-creation with ecosystem partners (Lu & Zhang, 2022).

The second key activity, termed **promoting the collaboration innovation process**, includes *building an innovation learning organization*. This initiative aims to develop key technical expertise and discover new applications for successful orchestration models (Leten et al., 2013). Importantly, it not only enhances the innovation potential of partners but also solidifies the leadership role of ecosystem orchestrators. In fostering innovative collaboration, orchestrators assume a dual role. First, they lead by introducing novel ideas and solutions to the ecosystem. Tian et al. (2021) suggest that orchestrators should introduce innovative problem-solving approaches to ecosystem partners and provide knowledge services to stimulate value co-creation. Second, orchestrators facilitate by *organizing agile innovation sprint meetings with partners* to foster the generation and exchange of innovative ideas (Masucci et al., 2020; Ritala et al., 2013). Internal learning within the organization and the recruitment of expertise can help bridge knowledge gaps (Lingens, Miehé, et al., 2021). Furthermore, collaboration with ecosystem partners can accelerate learning and innovation through the sharing and exchange of information, knowledge, and resources from various firms (Lingens & Huber, 2021; Poblete et al., 2022; Sjödin et al., 2022). Embracing a "fail fast, learn fast" mindset is crucial, along with continuously conducting training activities to thoroughly understand market needs (Marheine et al., 2021). Moreover, as Laczko et al. (2019) highlight, orchestrators actively involve partners in developing innovative products, thereby increasing the profitability and "stickiness" of partners to the ecosystem.

The third key activity, termed **commercializing and scaling innovative projects**, encompasses several critical steps. Initially, it involves *evaluating and configuring innovation project ideas* to capitalize on innovation opportunities within current partnerships (Aagaard & Rezac, 2022). During this phase, selection criteria, such as strategic importance, consistency, feasibility, urgency, and risk, are openly discussed and assessed to identify the most promising ideas (Gomes et al., 2021; Poblete et al., 2022). Following these evaluations, inter-organizational teams are swiftly formed to conduct preliminary studies and develop project plans. These form the basis for selecting innovative projects that merit further development (Poblete et al., 2022). The initial project proposal is then refined and expanded, culminating in the creation of a comprehensive project and risk plan (Reypens et al., 2021). It is crucial to note that innovative new products or services can only create value if successfully commercialized (Isckia et al., 2020). Additionally, high innovation potential significantly enhances a company's commercialization potential (Aagaard & Rezac, 2022). Orchestrators play a crucial role in *devising a commercialization roadmap for innovation solutions*, ensuring their commercial viability and scalability across global markets. This commercialization process requires efficient coupling and feedback mechanisms to ensure the commercial success of innovation (Isckia et al., 2020).

This leads to the following proposition:

**Proposition 5**. *In orchestrating ecosystems, it is recommended that orchestrators adopt ecosystem innovation practices, which include instigating ecosystem-centric innovation culture, promoting collaborative innovation process, and commercializing and scaling innovative projects. These activities facilitate a rapid response to market changes and promote first-mover advantages and competitive advantages.*

*4.6. Relationships among the ecosystem orchestration practices for industrial firms*

As depicted in Fig. 3, we present a synthesis of the interrelationships among five ecosystem orchestration practices: ecosystem strategic design practices, ecosystem relational practices, ecosystem resource integration practices, ecosystem technological leveraging practices, and ecosystem innovation practices. Taking the initial letters of each practice and emphasizing their continuous adjustment and dynamic nature with the "-ing" suffix, we have named this integrative conceptual framework the "**stirring model of ecosystem orchestration.**" These five practices metaphorically function as the stirrer of the ecosystem practices, maintaining dynamic equilibrium and preventing stagnation into a state of inertia. Firstly, the stirring model reflects the interactive characteristic of ecosystem orchestration practices with diverse outcomes. The interaction processes between these practices influence dual-directional effects. For example, ecosystem strategic design practices act as the guiding force, constantly shaping and directing the other four practices. Conversely, the other four practices reciprocally reinforce the actualization and success of ecosystem strategic design practices goals. Secondly, the dynamic model mirrors the ongoing iterative nature of ecosystem orchestration, facilitating adaptability to the organizational context and external dynamic changes. This iterative characteristic ensures the ecosystem's dynamics, fostering the attainment of sustainable competitive advantages.

Fig. 3 depicts a triangular relationship among ecosystem strategic design practices, ecosystem relational practices, and ecosystem resource integration practices. This configuration highlights the strategic design's guiding role and the fundamental importance of ecosystem relational and resource integration practices. Strategic design practices provide overarching direction for other orchestration activities. Moreover, ecosystem actors often engage in complex and intertwined relationships and resources, and the practices that orchestrate these elements lay a solid groundwork for overall ecosystem orchestration. Consequently, we propose:

**Proposition 6.1**. *Ecosystem strategic design practices, relational practices, and resource integration practices collectively constitute the core structure for effective ecosystem orchestration.*

The inner circle of Fig. 3, the other two practices, resembling two wings of a triangular orchestration structure, are constituted by innovation practices and technological leveraging practices. These practices play distinct roles in ecosystem orchestration for industrial firms. Firstly, they act as additional conduits linking ecosystem strategic design practices with ecosystem relational and resource integration practices. These connections enable orchestrators and partners to pursue a shared value proposition, expanding the value of resource and relationship orchestration practices. Advanced technologies and incremental innovations exert a leveraging effect in orchestrating resources and relationships. Deploying innovative technology can increase productivity and simultaneously decrease inter-firm communication costs. Incremental innovation practices facilitate the gradual enhancement of products or services and optimize resource distribution through experimentation and improvement. Secondly, technology and innovation serve as sources of relationships and resources. Disruptive technologies and radical innovation practices introduce digital resources, making attracting active partners and catalyzing new relationships and resources more feasible. Furthermore, as mentioned earlier, the interaction among these practices is bi-directional. Orchestrators and partners utilize ecosystem resources and the trust established in mutual relationships to continually advance technology and innovation (Aagaard & Rezac, 2022; Lingens, Huber, et al., 2022; Zeng et al., 2021). Additionally, reflecting the complex interplay of relationships and resources among partners, innovation, and technology are interdependent and tightly interwoven. Therefore, we propose:

**Proposition 6.2**. *Ecosystem technological leveraging practices and innovation practices are complementary ecosystem orchestration practices and play a crucial role in expanding ecosystem.*

In summary, the aforementioned practices constitute our stirring model of ecosystem orchestration. As depicted in the outer circle of Fig. 3, we assert that ecosystem strategic design practices play a pivotal role in forming a shared value proposition among partners, which is foundational for the ecosystem's existence. Driven by this shared value proposition, increasing partners and resources are continuously integrated into the ecosystem. Ecosystem relational practices enhance ecosystem stability, while ecosystem resource integration practices improve resource mobility. These practices contribute significantly to ecosystem resilience. Interdependent and reciprocal partnerships facilitate adaptability to market changes and risks, and the sharing and optimization of resources among ecosystem members promote sustainable resource utilization. The significance of technology and innovation is increasingly apparent in the digital era. Orchestrators leverage these elements for ecosystem transformation and breakthroughs. Ecosystem technological leveraging practices foster openness and reciprocity within the ecosystem, and ecosystem innovation practices enable the creation of first-mover advantages, contributing to ecosystem vitality. In essence, these five practices are all aligned toward achieving the joint value proposition of the ecosystem.

Therefore, we propose:

**Proposition 6.3**. *Ecosystem strategic design practices, relational practices, resource integration practices, technological leveraging practices, and innovation practices collectively form the "stirring model" of ecosystem orchestration, providing an competitive framework for orchestrating ecosystems.*

**5. Conclusion**

*5.1. Theoretical contributions*

In the era of Industry 4.0, where digital technologies rapidly evolve, ecosystems play a more crucial role than individual business actors. The contribution of this study to the ecosystem literature is twofold. First, in terms of methodology, we used a qualitative *meta*-analysis of ecosystem orchestration. This approach, which integrated 31 studies with 111 business cases, differs from previous ecosystem studies focusing on single industries or countries. By drawing insights from diverse companies, ranging from traditional manufacturing and service industries to digitally emerging high-tech companies, this study enhances the contextual applicability and external validity of our findings. Additionally, it enriches the literature with a broader, more diverse perspective on ecosystem orchestration, potentially leading to the emergence of new theories.

Secondly, from the perspective of the business actor in the ecosystem, our primary focus rests on the ecosystem orchestrator (i.e., focal firm, ecosystem leader). We systematically summarize the orchestration practices performed by the orchestrators, elucidating how they implement specific ecosystem orchestration activities. Our data analysis identifies five primary ecosystem orchestration practices (ecosystem strategic design practices, ecosystem relational practices, ecosystem resource integration practices, ecosystem technological leveraging practices, and ecosystem innovation practices) and delineates 15 key orchestration activities and 30 sub-orchestration activities. Moreover, we present a holistic framework that integrates these practices, responding to the recent call by Autio (2022) for research on specific actions that leading firms can take to orchestrate their ecosystem successfully.

*5.2. Managerial implications*

Given the diverse industries covered in our cases, our framework on ecosystem orchestration practices serves as a valuable practical guide for ecosystem leaders. More specifically, we offer insights for ecosystem orchestrators to develop specific roles and balance different orchestration practices. Especially beneficial for companies new to being orchestrators, our framework provides direction for effective orchestration. In addition, this paper details the specific orchestration activities that are operating in practice, offering a hands-on practical approach for ecosystem orchestrators and, thereby, minimizing costly errors and detours. Simultaneously, our findings assist experienced orchestrators in self-evaluation. By checking their operations against the 15 key orchestration activities, companies can identify areas for refinement, overcome development bottlenecks, and effectively realize common value propositions. In addition, companies that aspire to become ecosystem orchestrators can employ this framework to comprehensively evaluate their capabilities, enhance their strengths, and address their weaknesses, ultimately positioning themselves as orchestrators.

*5.3. Limitations and suggestions for future research*

The limitations of this paper reside in the research methodology and in the content of the study. Regarding research methodology, our qualitative *meta*-analysis approach presents certain challenges regarding replication. The complex case collection and filtering process can be daunting, and our study primarily relies on original authors' interpretations of raw qualitative data, potentially deviating from the raw data. Additionally, this process may have lost contextual information from the original studies (Combs et al., 2019). Nevertheless, our study provides a comprehensive and general understanding of ecosystem orchestration. Another limitation is the interpretation of the study topic. While we have carefully selected sources from multiple databases, there may still be omissions of important literature. Furthermore, our study did not explore the orchestration practices from a longitudinal perspective, potentially missing the dynamic changes that occur in different stages of ecosystem development.

Based on an analysis of the literature, we have identified several research gaps and proposed numerous avenues for future studies related to ecosystem orchestration practices. Table 4 presents the research gaps in the extant literature and the potential research questions.

**CRediT authorship contribution statement**

**Lei Shen:** Writing – original draft, Resources, Project administration, Investigation, Formal analysis, Data curation, Conceptualization. **Qingyue Shi:** Writing – original draft, Investigation, Data curation, Conceptualization. **Vinit Parida:** Writing – original draft, Supervision, Methodology, Conceptualization. **Marin Jovanovic:** Writing – review & editing, Supervision, Conceptualization.

**Funding**

This work was supported by the Ministry of Science and Technology of the People's Republic of China (No. G2021136006L).

**Declaration of competing interest**

The authors declare that they have no known competing financial

**Table 4**
Orchestration practices-based research gaps and questions.

| Ecosystem orchestration practices | Research gaps | Research questions |
|---|---|---|
| Ecosystem strategic design practices | ● Evolution of value proposition design  Formulation of ecosystem identity  Process for evaluating market | 1. How do ecosystem orchestrators continuously innovate the design of shared value propositions to meet the requirements of ecosystem evolution?  What are the key elements employed by orchestrators to establish a distinctive ecosystem identity?  How does the identity of the orchestrator influence ecosystem identity legitimation?  What factors influence the acceptance and adoption of ecosystem identity by complementary partners?  What key indicators should orchestrators prioritize when scanning new markets? |
| Ecosystem relational practices | ● The role of ecosystem partners  Paradox of partner diversity  Degree of empowerment | 1. What role do ecosystem partners play in implementing the ecosystem orchestration process?  What relational factors influence partner engagement and commitment?  What opportunities and challenges does partner diversity impose on ecosystem orchestration?  How can the ecosystem orchestrator reconcile the dual imperatives of empowerment and regulation?  How does the empowerment of partners lead to achieving optimal outcomes for the ecosystem? |
| Ecosystem resource integrating practices | ● Dynamic resource integration  The security and privacy of resources  Benefit from distribution of resources | 1. How can orchestrators adapt resource integration practices in response to changing ecosystem dynamics?  What key ecosystem characteristics attract investment from external investors?  What strategies do orchestrators utilize to secure funding from external investors?  How can orchestrators guarantee the security and privacy of shared resources, especially in contexts involving sensitive or proprietary data?  How are the benefits generated by shared resources distributed among ecosystem partners? |
| Ecosystem technological leveraging practices | ● Cost-benefit analysis of building digital infrastructure  Data flow between partners | 1. How should a firm choose an appropriate revenue model given the digital infrastructure development and maintenance costs involved? |

(*continued on next page*)

*L. Shen et al.**Journal of Business Research 173 (2024) 114463*

**Table 4** (*continued*)

| Ecosystem orchestration practices | Research gaps | Research questions |
|---|---|---|
| | Accelerating effect of technology | How do the technological capabilities of the orchestrator affect ecosystem development? What are the risks and challenges with inter-organizational data flows, and how do orchestrators address them? How can technology exert an augmented impact on the orchestration process? |
| Ecosystem innovation practices | • The uniqueness of innovation-centric ecosystem culture  Collaborative innovation mechanism  Innovation project operation | 1. What are the underlying characteristics of innovation-centric culture in ecosystem design and development?  What factors influence innovative collaborative processes, and how do they unfold?  What are the key challenges faced by orchestrators in maintaining an innovation learning environment?  How does the ecosystem orchestrator evaluate innovation project ideas from ecosystem partners?  What is the distribution pattern of benefits generated by innovation projects among different ecosystem partners? |

interests or personal relationships that could have appeared to influence the work reported in this paper.

**Lei Shen** is a Professor in Glorious Sun School of Business & Management at Donghua University, China. Her research interests include intelligent commerce, marketing, and organizational changing. She has published numerous journal articles, including articles in Technological Forecasting and Social Change, Industrial Marketing Management, Technovation, and others.

**Qingyue Shi** is a doctoral student at Glorious Sun School of Business & Management at Donghua University. Her research interests include ecosystems, orchestration capability, inter-organizational cooperation, and strategic alliances.

**Vinit Parida** is Professor of Entrepreneurship and Innovation at Luleå University of Technology, Sweden, and a visiting professor at the University of Vaasa, Finland. His research interests include organizational capabilities, servitization, business model innovation, digitalization of industrial ecosystems, and the circular economy. He has published numerous journal articles, including articles in Strategic Management Journal, Industrial Marketing Management, Journal of Management Studies, Production and Operations Management, Journal of Cleaner Production, and he is an editorial review board member for Industrial Marketing Management and Journal of Business Research. He is the recipient of multiple awards for his research.

**Marin Jovanovic** is an associate professor at the department of operations management at Copenhagen Business School. His research has been published in academic journals, such as Organization Science, Journal of Product Innovation Management, R&D Management, International Journal of Operations & Production Management, Technovation, International Journal of Production Economics, Journal of Business Research, and Industrial Marketing Management. His research interests include the digital transformation of manufacturing, maritime, and healthcare sectors, platform ecosystems in the business-to-business context, and artificial intelligence. Marin has held positions at the ESADE Business School and the University of Cambridge.